\documentclass[12pt]{article}

\begin{document}

\title{\textbf{Quantum Corrections for a Bardeen Regular Black Hole}}

\author{M. Sharif \thanks{msharif@math.pu.edu.pk} and Wajiha Javed \\
\\Department of Mathematics, University of the Punjab,\\Quaid-e-Azam
Campus, Lahore-54590, Pakistan.}
\date{}

\maketitle

In this paper, we study the quantum corrections to the
thermodynamical quantities (temperature and entropy) for a Bardeen
charged regular black hole by using a quantum tunneling approach
over semiclassical approximations. Taking into account the quantum
effects, the semiclassical Bekenstein-Hawking temperature and the
area law are obtained, which are then used in the first law of
thermodynamics to evaluate corrections to these quantities. It is
interesting to mention here that these corrections reduce to the
corresponding corrections for the Schwarzschild black hole when
the charge $e=0$.\\


\section*{I. INTRODUCTION}

According to General Relativity, a black hole (BH) is a region of
space from which nothing, including light, can escape. Quantum
mechanics suggests that BHs are not black, but possess temperature
and emit radiation (energy) continuously, called the Hawking
radiation [1]. Bekenstein [2] predicted that BHs
should have finite, non-zero temperature and entropy. Many
attempts [3, 4] have been made to address the quantum
mechanics of a scalar particle to obtain BH radiation. The entropy
of the Kerr-Newmann and the de Sitter spacetimes has been found to
be always equal to one quarter the area of the event horizon in
fundamental units. However, tunneling provides the best way to
visualize the source of radiation as compared to the approach
mentioned above. Quantum tunneling refers to the phenomenon of
particle's ability to penetrate energy barriers within electronic
structures. This technique is based on electron-positron pair
production, which requires an electric field.

It is a well-known fact that as a particle crosses the event
horizon, there is a change of energy. For tunneling, it is assumed
that the particles follow trajectories that are not allowed
classically [5-7]. The evaporation of a BH is
related to the emission of quantum particles as Hawking radiation
allows BHs to lose mass. Black holes that lose more matter than
they gain through other means are expected to dissipate, shrink
and ultimately vanish. This process causes a change in the
characteristic parameters (mass, angular momentum and charge),
which alters the thermodynamics of BH. The motion of the particle
may be in the form of outgoing or ingoing radial null geodesics.
For the outgoing geodesics, the particle must be imaginary while
for the ingoing geodesics, it is assumed to be real. This is due
to the fact that only a real particle that has a speed less than
or equal to the speed of light can exist inside the event horizon.

In order to calculate the imaginary part of the action, Parikh and
his collaborator [5] used a procedure based on the purely radial
null geodesic method. Srinivasan and Padmanabhan [6] introduced
another technique based on the Hamilton-Jacobi method. Planck's
scale correction in the Parikh-Wilczek tunneling framework was
studied by many people [8-10]. Jiang \emph{et al.} [8] investigated
Hawking radiation as massless charged particles tunneling across the
event horizon of Kerr and Kerr-Newmann BHs. Xu and Chen [9]
evaluated the total flux of Hawking radiation for the Kerr-(anti)
de-Sitter BHs. Liu and Zhu [10] found the emission rate of massless
particles tunneling through the corrected horizon. Banerjee and
Majhi [11] discussed the role of chirality by connecting the anomaly
and the tunneling formalisms for Hawking radiation from BH. The same
authors [12] also obtained the Hawking black body spectrum with the
appropriate temperature for a BH. Majhi [13] derived the Hawking
radiation and performed BH thermodynamic spectroscopy; i.e., the
spectrum of entropy and area were obtained by using a density matrix
technique. Banerjee \emph{et al.} [14] found the entropy spectrum of
a BH by using the tunneling criterion.

The quantum geometry of the BH horizon has been studied using loop
quantum gravity. Loop quantization reproduces the result of BH
entropy originally discovered by Bekenstein and Hawking. Further,
it leads to the computation of a quantum gravity correction to the
entropy and the radiation of a BH.

Banerjee and Majhi [7] analyzed Hawking radiation as tunneling by
using the Hamilton-Jacobi method. They computed quantum corrections
to the Hawking temperature and the Bekenstein-Hawking area law of
the Schwarzschild, anti-de Sitter Schwarzschild and Kerr BHs.
Banerjee and Modak [15] gave a new conceptually simple approach to
obtain the entropy for any stationary BH. They determined the
semiclassical BH entropy for the most general Kerr-Newmann
spacetime. Following the same approach, Akbar and Saifullah [16]
studied quantum corrections to the entropy and the horizon area for
the Kerr-Newmann, charged rotating BTZ and Einstein-Maxwell
dilaton-axion BHs. Recently, Larra$\tilde{n}$aga [17] extended this
type of work for a charged BH of string theory and for the Kerr-Sen
BH. In this paper, we work out the temperature and the entropy
corrections for a Bardeen regular BH. This is a generalization of
the entropy correction for the Schwarzschild BH [7].

Banerjee and Majhi [18] computed corrections to the Hawking
temperature and Bekenstein-Hawking entropy for the Schwarzschild BH
by using the tunneling formalism based on the quantum WKB
approximation. Majhi [19] analyzed the Hawking radiation as
tunneling of a Dirac particle through an event horizon by applying
the Hamilton-Jacobi method beyond the semiclassical approximation.
Majhi and Samanta [20] investigated the tunneling of a photon and a
gravitino through an event horizon by applying the Hamilton-Jacobi
method beyond the semiclassical approximation. Banerjee \emph{et
al.} [21] discussed the quantum gravitational correction to the
Hawking temperature from the Lemaitre-Tolman-Bondi model in a
semiclassical approximation. They obtained the standard expression
for the Hawking temperature and its first quantum gravitational
correction.

The plan of this paper is as follows: Section \textbf{II} describes
basic equations for the corrections using the Hamilton-Jacobi
method, the first law of thermodynamics and the exactness condition.
In Section \textbf{III}, we evaluate semiclassical thermodynamical
quantities for a Bardeen regular BH. These quantities evaluate
corrections to the temperature and the entropy. Finally, in the last
section, we present the outlook of the paper.

\section*{II. REVIEW}

In this section, we review some basic material used to evaluate
the corrections to the entropy and the temperature. We use the
Hamilton-Jacobi method [7] to compute the imaginary part of
the action outside the semiclassical approximation by admitting
all possible quantum corrections. Here, we write the expression
for the quantum correction of a general function $S(r,t)$. We
expand this function in terms of a series in powers of $\hbar$,
\emph{i.e.},
\begin{eqnarray}
S(r,t)=S_0(r,t)+\hbar S_1(r,t)+\hbar^2 S_2(r,t)+....
=S_0(r,t)+\sum_{i}\hbar^i S_i(r,t),  \label{1}
\end{eqnarray}
where $i=1,2,3,....,$ $S_0$ is the semiclassical value and the
terms involving $\hbar$ and its higher powers are considered as
correction terms. The dimension of $S_i$ (proportional to $S_0$)
is $\hbar$ while the proportionality constants is
$(\hbar^i)^{-1}$. The Planck's constant, $\hbar$, is of the order
of the square of the Planck's mass (for $G=c=1$). Dimensional
analysis provides the proportionality constants with dimensions of
$m^{-2i}$, where $m$ is the mass of the BH. The most general
expression for $S$ in Eq. (\ref{1}) can be written as
\begin{eqnarray}
S(r,t)=S_0(r,t)+\sum_{i}\alpha_i\frac{\hbar^i}{m^{2i}}S_0(r,t)
=S_0(r,t)\left(1+\sum_{i}\alpha_i\frac{\hbar^i}{m^{2i}}\right),\label{2}
\end{eqnarray}
where $S_0(r,t)$ is the semiclassical entropy and the remaining
terms represent quantum corrections.

The modified form of the temperature of the BH can be written as
\begin{equation}
T=T_H{\left(1+\sum_{i}\alpha_i\frac{\hbar^i}{m^{2i}}\right)}^{-1},\label{11111}
\end{equation}
where $T_H$ is the standard semiclassical Hawking temperature and
the terms with $\alpha_i$ are corrections due to quantum effects
[7]. The dimensionless parameter $\alpha_i$ corresponds to
the higher order loop corrections to the surface gravity
$\kappa=2\pi T$ of the BH. The corrected form of surface gravity is
given by
\begin{equation}
\kappa=\kappa_0{\left(1+\sum_{i}\alpha_i\frac{\hbar^i}{m^{2i}}\right)}^{-1},\label{loppp}
\end{equation}
where $\kappa_0=2\pi T_H$ is the standard semiclassical surface
gravity. If we consider $\alpha_i$ in terms of a single
dimensionless parameter $\beta$ such that $\alpha_i=\beta^i$, then
we get
\begin{equation}
1+\sum_{i}\alpha_i\frac{\hbar^i}{m^{2i}}=1+\left(\frac{\beta\hbar}{m^2}+\frac{\beta^2\hbar^2}{m^4}
+\frac{\beta^3\hbar^3}{m^6}+
\frac{\beta^4\hbar^4}{m^8}+...\right)=\left(1-\frac{\beta\hbar}{m^2}\right)^{-1}.\label{}
\end{equation}
Consequently, Eq. (\ref{loppp}) reduces to
\begin{equation}
\kappa=\kappa_0\left(1-\frac{\beta\hbar}{m^2}\right);\label{}
\end{equation}
hence, Eq. (\ref{11111}) can be written as
\begin{equation}
T=T_H\left(1-\frac{\beta\hbar}{m^2}\right).\label{22222}
\end{equation}

Now we follow Ref.15 to evaluate an expression of the entropy for a
charged regular BH. The first law of thermodynamics for two
parameters $m$ and $e$, \emph{i.e.}, the mass and charge of the BH,
respectively, can be written as
\begin{equation}
dm=TdS+\Phi de,\label{4}
\end{equation}
where $T,~S$ and $\Phi$ are the temperature, entropy and
electrostatic potential of the BH, respectively. Equation
(\ref{4}) can be written as
\begin{equation}
dS(m,e)=\frac{1}{T}dm-\frac{\Phi}{T}de\label{5}.
\end{equation}
A differential of a function $f=f(x,y)$ is given by
\begin{equation}
df(x,y)=A(x,y)dx+B(x,y)dy,\label{6}
\end{equation}
which is an exact differential if the following conditions hold
\begin{eqnarray}
\frac{\partial A}{\partial y}&=&\frac{\partial B}{\partial x };\\
\frac{\partial f}{\partial x}&=& A,\\ \frac{\partial f}{\partial
y}&=&B.\label{7}
\end{eqnarray}
It follows from Eq. (\ref{6}) that
\begin{equation}
f(x,y)=\int A dx+\int B dy-\int\left(\frac{\partial}{\partial y
}\left(\int A dx\right)\right)dy.\label{9}
\end{equation}
Comparing Eqs. (\ref{5}) and (\ref{6}), we get $A=\frac{1}{T}$ and
$B=-\frac{\Phi}{T}$, where $m$ and $e$ will play the roles of $x$
and $y$, respectively. The condition for an exact differential
will become
\begin{equation}
\frac{\partial}{\partial e}\left(\frac{1}{T}\right)=
\frac{\partial}{\partial m}\left(-\frac{\Phi}{T}\right).\label{10}
\end{equation}
Using Eq. (\ref{9}), we can write the entropy for the BH in the
integral form as
\begin{equation}
S(m,e)=\int\frac{1}{T}dm-\int\frac{\Phi}{T}de-\int\left(\frac{\partial}{\partial
e}\left(\int\frac{1}{T}dm\right)\right)de.\label{11}
\end{equation}

When the thermodynamical quantities satisfy Eqs. (\ref{4}) and
(\ref{10}), $dS$ will be an exact differential. We can use the
integral form in Eq. (\ref{11}) to work out the semiclassical
entropy for this black hole. It follows from Eq. (\ref{10}) that
the quantities $T$ and $\Phi$ satisfy
\begin{equation}
\frac{\partial}{\partial
e}\int\left(\frac{dm}{T}\right)=-\frac{\Phi}{T}.\label{12}
\end{equation}
Using this equation in Eq. (\ref{11}), we obtain
\begin{equation}
S(m,e)=\int\frac{1}{T}dm.\label{13}
\end{equation}
This equation shows the semiclassical entropy of the charged BH,
which is independent of $\Phi$.

\begin{center}
\section*{III. THERMODYNAMICAL QUANTITIES FOR THE BARDEEN MODEL}
\end{center}

When particles escape, a BH loses a small amount of its energy
(mass). The power emitted by a BH in the form of Hawking radiation
can be estimated for a charged regular BH. The generalized form of
BH solutions is given by
\begin{equation}
{ds}^2=-F{dt}^2+F^{-1}{dr}^2+r^2{d\theta}^2+r^2\sin^2\theta{d\phi}^2,\label{14}
\end{equation}
where $F=1-2\frac{M(r)}{r}$. This metric can be reduced to a
well-known BH for a special choice of $M(r)$. Ayon-Beato and Gracia
[22] gave a physical interpretation of the Bardeen model [23] by
showing that the charge associated with it acts as a magnetic
monopole charge. This is described by the metric with
\begin{equation}
M(r)=\frac{m{r^3}}{(r^2+e^2)^\frac{3}{2}}.\label{15}
\end{equation}
Here, $m$ and $e$ stand for the mass and the monopole charge of a
self-gravitating magnetic field of a non-linear electrodynamics
source, respectively. This solution exhibits BH behavior for
$e^2\leq(16/27)m^2$ while it reduces to the Schwarzschild solution
for $e=0$.

The Bardeen regular black hole solution has a spherical event
horizon at
\begin{equation}
r_+=2M(r_+),\label{23}
\end{equation}
where $r_+$ is the event horizon. Replacing the value of $M$, it
follows that
\begin{equation}
1-\frac{2m{r_+^2}}{(r_+^2+e^2)^{\frac{3}{2}}}=0,\label{16}
\end{equation}
whose roots are given in Ref.16 while its area [17] is
\begin{equation}
A=\int\sqrt{g_{\theta\theta}g_{\varphi\varphi}}d\theta\varphi=4\pi
r_+^2.\label{24}
\end{equation}
Let us write
\begin{equation}
F(r)=1-\frac{2m{r^2}}{(r^2+e^2)^{\frac{3}{2}}}.\label{17}
\end{equation}
The event horizon related with temperature $T_H$ [24, 25] is
\begin{equation}
T_H={\frac{\hbar F'(r)}{4\pi}}|_{r=r_+}=\frac{\hbar m r_+(r_+^2-2
e^2)}{2\pi(r_+^2+e^2)^{\frac{5}{2}}}, \label{18}
\end{equation}
where $F'(r)$ denotes the derivative of $F$ with respect to $r$ and
$m=\frac{(r^2+e^2)^\frac{3}{2}}{2{r^2}}$. The electric potential is
given by [26]
\begin{equation}
\Phi=\frac{\partial m}{\partial e}|_{r=r_+}=\frac{3 e}{2
r_+^2}(r_+^2+e^2)^\frac{1}{2}.\label{19}
\end{equation}

With these thermodynamical quantities, the Bardeen regular BH
satisfies the first law of thermodynamics, Eq. (\ref{4}), and the
condition in Eq. (\ref{10}). Thus, the semiclassical entropy takes
the form
\begin{equation}
S_0(m,e)=\int\frac{1}{T_H}dm=\frac{2\pi}{\hbar}\int\frac{(r_+^2+e^2)^\frac{5}{2}}{m
r_+(r_+^2-2e^2)}dm.\label{20}
\end{equation}
To evaluate this integral, we use Eq. (\ref{16}), which yields
\begin{equation}
dm=\frac{(r_+^2+e^2)^\frac{1}{2}(r_+^2-2e^2)}{2
r_+^3}dr_+.\label{21}
\end{equation}
Using this value in Eq. (\ref{20}), we obtain semiclassical entropy
\begin{equation}
S_0=\frac{2\pi}{\hbar}\int
r_+(1+\frac{e^2}{r_+^2})^\frac{3}{2}dr_+=\frac{1}{2\hbar}\int\frac{A}{r_+}
(1+\frac{e^2}{r_+^2})^\frac{3}{2}dr_+.\label{22}
\end{equation}
The integrated form of the above expression is
\begin{eqnarray}
S_0&=&2\pi\hbar^{-1}\left((-\frac{e^2}{r}+\frac{r}{2})\sqrt{e^2+r^2}+
\frac{3}{2}{e^2}\ln(r+\sqrt{e^2+r^2})\right).
\end{eqnarray}
Note that if we put charge $e=0$ and $\hbar=1$, we recover the
Bekenstein-Hawking area law relating entropy and the horizon area,
$S_0=\frac{A}{4}$, as usually occurs in the Einstein gravity. In the
following, we work out the corrected form by taking into account the
quantum effects on the thermodynamical quantities (temperature and
entropy) inside the event horizon of the charged regular BH.

\subsection*{1. Hawking Temperature Corrections}

In this section, we find the correction to the Hawking temperature
as a result of quantum effects for the Bardeen regular BH. The
expression for the semiclassical Hawking temperature, Eq.
(\ref{18}), turns out to be
\begin{equation}
T_H=\frac{\hbar(r_+^2-2 e^2)}{4\pi r_+(r_+^2+e^2)}. \label{lmjo}
\end{equation}
The corrected temperature, Eq. (\ref{22222}), in terms of the
horizon radius can be written as
\begin{equation}
T=T_H\left(1-\frac{4\beta\hbar{r^4}}{(r^2+e^2)^3}\right).
\label{tiol}
\end{equation}
Using Eq. (\ref{tiol}) in Eq. (\ref{lmjo}), we obtain the quantum
correction of temperature $T$ as
\begin{equation}
T=\frac{\hbar(1-2\frac{e^2}{r_+^2})}{4\pi
r_+(1+\frac{e^2}{r_+^2})}\left(1-\frac{4\hbar\beta}{r_+^2}
(1+\frac{e^2}{r_+^2})^{-3}\right). \label{12345}
\end{equation}
For $e=0$, this reduces to the modified Hawking temperature of the
Schwarzschild with
\begin{equation}
\beta=-\frac{1}{360\pi}\left(-N_0-\frac{7}{4}N_{\frac{1}{2}}
+13 N_1+\frac{233}{4}N_{\frac{3}{2}}-212 N_2\right),
\end{equation}
where $N_s$ refers to the number of spin '$s$' fields [7].

\subsection*{2. Entropy Corrections}

Here, we evaluate the quantum corrections to the entropy of the
Bardeen charged regular BH. In terms of the horizon radius, the
corrected form of entropy, Eq. (\ref{2}), is given by
\begin{equation}
S(r,t)=S_0(r,t) \left(
1+\sum_{i}\frac{\alpha_i\hbar^i({4r_+^4})^i}{(r_+^2+e^2)^{3i}}
\right) \label{3}
\end{equation}
while the corrected form of the Hawking temperature can be written
as
\begin{equation}
T=T_H{\left(1+\sum_{i}\frac{\alpha_i\hbar^i({4r_+^4})^i}{(r_+^2+e^2)^{3i}}\right)}^{-1}.\label{}
\end{equation}
In the first law of thermodynamics, Eq. (\ref{5}), we replace the
temperature $T$ by the corrected form of the temperature.
Similarly, due to the corrected temperature, Eq. (\ref{10}) takes
the following form:
\begin{equation}
\frac{\partial}{\partial
e}\left(\frac{1}{T_H}\right){\left(1+\sum_{i}
\frac{\alpha_i\hbar^i({4r_+^4})^i}{(r_+^2+e^2)^{3i}}\right)}=
\frac{\partial}{\partial
m}\left(-\frac{\Phi}{T_H}\right){\left(1+\sum_{i}
\frac{\alpha_i\hbar^i({4r_+^4})^i}{(r_+^2+e^2)^{3i}}\right)}\label{}.
\end{equation}
Thus, the entropy with the correction terms is given by
\begin{eqnarray}
S(m,e)&=&\int\frac{1}{T_H}{\left(1+\sum_{i}
\frac{\alpha_i\hbar^i({4r_+^4})^i}{(r_+^2+e^2)^{3i}}
\right)}dm-\int\frac{\Phi}{T_H}{\left(1+\sum_{i}
\frac{\alpha_i\hbar^i({4r_+^4})^i}{(r_+^2+e^2)^{3i}}
\right)}de\nonumber\\
&-&\int\left(\frac{\partial}{\partial
e}\left(\int\frac{1}{T_H}{\left(1+\sum_{i}\frac{\alpha_i\hbar^i({4r_+^4})^i}{(r_+^2+e^2)^{3i}}
\right)}dm\right)\right)de.\label{abc}
\end{eqnarray}
This is the corrected and modified form of Eq. (\ref{11}).

We can simplify these complicated integrals by employing the
exactness criterion described above. As a result, Eq. (\ref{abc})
reduces to
\begin{equation}
S(m,e)=\int\frac{1}{T_H}{\left(1+\sum_{i}
\frac{\alpha_i\hbar^i({4r_+^4})^i}{(r_+^2+e^2)^{3i}}
\right)}dm,\label{}
\end{equation}
which can be written in expanded form as
\begin{eqnarray}
S(m,e)&=&\int\frac{1}{T_H}dm+\int{\frac{\alpha_1\hbar({4r_+^4})}{T_H(r_+^2+e^2)^3}}dm+
\int{\frac{\alpha_2\hbar^2({4r_+^4})^2}{T_H(r_+^2+e^2)^6}}dm\nonumber\\
&+&\int{\frac{\alpha_3\hbar^3({4r_+^4})^3}{T_H(r_+^2+e^2)^9}}dm+....\nonumber\\
&=&I_1+I_2+I_3+I_4+...,
\end{eqnarray}
where the first integral $I_1$ has been evaluated in Eq.
(\ref{22}) and $I_2, I_3,...$ are quantum corrections. Thus,
\begin{equation}
I_2=2^3\pi\alpha_1\int\frac{r_+^2}{(r_+^2+e^2)^\frac{3}{2}}dr_+
\label{}
\end{equation}
and
\begin{equation}
I_3=2^5\pi\alpha_2\hbar\int\frac{r_+^6}{(r_+^2+e^2)^\frac{9}{2}}dr_+.
\label{}
\end{equation}
In general, we can write
\begin{equation}
I_k=2^{2k-1}\pi\alpha_{k-1}\hbar^{k-2}
\int\frac{r_+^{4k-6}}{(r_+^2+e^2)^\frac{6k-9}{2}}dr_+,\quad k>3.
\label{}
\end{equation}
Therefore, the entropy with quantum corrections is given by
\begin{eqnarray}
S(m,e)&=&2\pi\hbar^{-1}\int\frac{(r_+^2+e^2)^\frac{3}{2}}{r_+^2}dr_++
2^3\pi\alpha_1\int\frac{r_+^2}{(r_+^2+e^2)^\frac{3}{2}}dr_+\nonumber\\&+&\sum_{k>2}
2^{2k-1}\pi\alpha_{k-1}\hbar^{k-2}\int\frac{r_+^{4k-6}}{(r_+^2+e^2)^\frac{6k-9}{2}}dr_+.
\label{123asbc}
\end{eqnarray}
This gives the quantum correction to the entropy for a Bardeen
charged BH.

For $e=0$, Eq. (\ref{123asbc}) reduces to the entropy correction
of the Schwarzschild black hole [7]; \emph{i.e.},
\begin{equation}
S=\frac{A}{4\hbar}+4\pi\alpha_1{\ln
A}-\frac{64\pi^2\hbar\alpha_2}{A}+...,\label{efg}
\end{equation}
where the area of the horizon, $A$, is given by Eq. (\ref{24}). It
is worth mentioning here that the first term of Eq. (\ref{efg}) is
the semiclassical Bekenstein-Hawking area law, \emph{i.e.},
$S_{BH}=\frac{A}{4\hbar}$; other terms are quantum corrections.
Thus, $S_{BH}$ is modified by quantum effects. After the integrals
are evaluated, Eq. (\ref{123asbc}) takes the form
\begin{eqnarray}
S(m,e)&=&2\pi\hbar^{-1}\left((-\frac{e^2}{r}+\frac{r}{2})\sqrt{e^2+r^2}+
\frac{3}{2}{e^2}\ln(r+\sqrt{e^2+r^2})\right)\nonumber\\
&&+ 2^3\pi\alpha_1\left(\frac{-r}{\sqrt{e^2+r^2}}
+\ln(r+\sqrt{e^2+r^2})\right)\nonumber\\
&&+{2^5}\pi\hbar{\alpha_2}
\left(\frac{r^7}{7{e^2}({e^2+r^2})^{\frac{7}{2}}}\right)+....
\label{123456}
\end{eqnarray}
The entropy in Eq. (\ref{123asbc}) in terms of $A$ is given as
follows:
\begin{eqnarray}
S(m,e)&=&\pi\sqrt{4\pi}\hbar^{-1}\int\frac{(\frac{A}{4\pi}+e^2)^\frac{3}{2}}{A^\frac{3}{2}}dA+
\frac{\alpha_1}{\sqrt{4\pi}}
\int\frac{\sqrt{A}}{(\frac{A}{4\pi}+e^2)^\frac{3}{2}}dA\nonumber\\
&+&\sum_{k>2}\frac{2^{2k-4}\hbar^{k-2}\alpha_{k-1}}{(4\pi)^\frac{4k-7}{2}}
\int\frac{A^\frac{4k-7}{2}}{(\frac{A}{4\pi}+e^2)^\frac{6k-9}{2}}dA.\label{erdsa}
\end{eqnarray}
When we take $e=0$, this equation leads to Eq.(\ref{efg}). Solving
Eq. (\ref{erdsa}), we obtain
\begin{eqnarray}
&&S(m,e)=\frac{\pi\sqrt{4\pi}\hbar^{-1}}{8\sqrt{A}\pi^{\frac{3}{2}}}\nonumber\\
&&\left((A-8\pi e^2)\sqrt{A+4\pi e^2}+12\sqrt{A}e^2\pi\ln
(\frac{\sqrt{\pi}}{2}(\sqrt{A}+\sqrt{A+4\pi e^2}))\right)\nonumber\\
&+&\frac{\alpha_1 16\pi^{\frac{3}{2}}}{\sqrt{4\pi}}
\left(\frac{-\sqrt{A}}{\sqrt{A+4e^2\pi}}
+\ln(\frac{\sqrt{\pi}}{2}(\sqrt{A}+\sqrt{A+4\pi e^2}))\right)\nonumber\\
&+&4\hbar\alpha_2\left(\frac{8\pi A^{\frac{7}{2}}}{7 e^2(A+4
e^2\pi)^{\frac{7}{2}}}\right)+....\label{rtyu}
\end{eqnarray}

\begin{center}
\section*{IV. OUTLOOK}
\end{center}

Black holes are sites of immense gravitational attraction into
which surrounding matter is drawn by gravitational forces.
Classically, the gravitation is so powerful that nothing, not even
electromagnetic radiation, can escape from the BH. Hawking defined
a BH as a galactic monster that emits radiation due to quantum
effects. The physical interpretation of this emission process
indicates that vacuum fluctuations accelerate particle (positive
mass)-antiparticle (negative mass) pairs towards the event horizon
of the BH. Hawking realized that a particle with positive mass has
enough energy to escape from the BH while a particle with negative
mass has no capability to escape from the BH and, hence, would
fall in. This process of an in-falling negative energy particle
eventually causes the mass of the BH to decrease. However, a
particle that goes off to a distant observer would be observed as
heat radiation. In fact, this process is a quantum tunneling
effect, in which a particle-antiparticle pair will form from the
vacuum, and one with positive energy will tunnel outside the event
horizon with a complex action that appears as Hawking radiation.
These radiations depend on the mass, angular momentum and charge
of the BH. Negative energy particles are the main source of
evaporation of a BH.

Using the above analysis, we have studied the quantum corrections
to the temperature and the entropy of the Bardeen regular BH. The
quantum correction to temperature is given by Eq. (\ref{12345}),
which reduces to the Schwarzschild temperature [7] for
$e=0$. We write down the first law of thermodynamics for BHs as a
differential of the entropy, including two parameters, mass and
charge. Applying the condition for exactness of differentials, we
evaluated the corrected entropy as a power series. For the charge
to be zero, Eq. (\ref{123asbc}) yields the corrected entropy of
the Schwarzschild BH (\ref{efg}). Here, the first term is the
semiclassical value while the leading correction term is
logarithmic. The other terms involve ascending powers of the
inverse of the area [7]. The entropy correction in terms of
the horizon area (\ref{erdsa}) also reduces to the Schwarzschild
BH for the charge to be zero.

\vspace{0.25cm}

\begin{center}
{\bf ACKNOWLEDGMENT}
\end{center}

\vspace{0.25cm}

We would like to thank the Higher Education Commission, Islamabad,
Pakistan, for its financial support through the {\it Indigenous
Ph.D. 5000 Fellowship Program Batch-IV}.

\begin{center}
{\bf \large REFERENCES}
\end{center}

\begin{description}

\item{[1]} S. W. Hawking, Nature {\bf 248}, 30 (1974).

\item{[2]} J. D. Bekenstein, Nuovo Cimento Lett. {\bf 4}, 737 (1972).

\item{[3]} J. B. Hartle and S. W. Hawking, Phys. Rev. D {\bf 13}, 2188 (1976).

\item{[4]} G. W. Gibbons and S. W. Hawking, Phys. Rev. D {\bf 15}, 2752 (1977).

\item{[5]} M. K. Parikh and F. Wilczek, Phys. Rev. Lett. {\bf 85}, 5042
(2000);\\
M. K. Parikh, Gen. Relativ. Gravit. \textbf{36}, 2419
(2004) [Int. J. Mod. Phys. D \textbf{13}, 2351 (2004)].

\item{[6]} K. Srinivasan and T. Padmanabhan, Phys. Rev. D {\bf 60}, 024007 (1999).

\item{[7]} R. Banerjee and B. R. Majhi, J. High Energy Phys. {\bf 06}, 095 (2008).

\item{[8]} Q-Q. Jiang, S-Q. Wu  and X. Cai, Phys. Rev. D {\bf 73}, 064003 (2006).

\item{[9]} Z. Xu and B. Chen, Phys. Rev. D {\bf 75}, 024041 (2007).

\item{[10]} C-Z. Liu and J-Y. Zhu, Gen. Relativ. Gravit. \textbf{40}, 1899 (2008).

\item{[11]} R. Banerjee and B. R. Majhi, Phys. Rev. D {\bf 79}, 064024 (2009).

\item{[12]} R. Banerjee and B. R. Majhi, Phys. Lett. B {\bf 675}, 243 (2009).

\item{[13]} B. R. Majhi, Phys. Lett. B {\bf 686}, 49 (2010).

\item{[14]} R. Banerjee, B. R. Majhi and E. C. Vagenas, Phys. Lett. B {\bf 686}, 279 (2010).

\item{[15]} R. Banerjee and S. K. Modak, J. High Energy Phys. {\bf 0905}, 063 (2009).

\item{[16]} M. Akbar and K. Saifullah, Eur. Phys. J. C (2010, in press), gr-qc/1002.3581;
gr-qc/1002.3901.

\item{[17]} A. Larra$\tilde{n}$aga, gr-qc/1003.2383; gr-qc/1003.2973.

\item{[18]} R. Banerjee and B. R. Majhi, Phys. Lett. B {\bf 674}, 218 (2009).

\item{[19]} B. R. Majhi, Phys. Rev. D {\bf 79}, 044005 (2009).

\item{[20]} B. R. Majhi and S. Samanta, Annals Phys. (2010, to appear); gr-qc/0901.2258.

\item{[21]} R. Banerjee, C. Kiefer and B. R. Majhi, gr-qc/1005.2264.

\item{[22]} E. Ay$\acute{o}$n-Beato and A. Garcia, Phys. Lett. B {\bf 493}, 149 (2000).

\item{[23]} J. Bardeen, \emph{Proceedings of GR5} (Tiflis, USSR, 1968).

\item{[24]} D. Kothawala, S. Sarkar and T. Padmanabhan, Phys. Lett. B {\bf 652}, 338 (2007).

\item{[25]} M. Akbar, Chin. Phys. Lett. {\bf 24}, 1158 (2007).

\item{[26]} M. Akbar and A. A. Siddiqui, Phys. Lett. B {\bf 656}, 217 (2007).

\end{description}
\end{document}